\documentclass[12pt]{JHEP3}
\usepackage{mathrsfs}
\usepackage{amsmath,amssymb}
\usepackage{epsfig}
\usepackage{relsize}
\usepackage{graphicx}
\input epsf

\def\x'{\mathaccent 19 x}
\def\y'{\mathaccent 19 y}
\def\n'{\mathaccent 19 n}
\def\u'{\mathaccent 19 u}

\def\et'{\mathaccent 19 \eta}
\def\th'{\mathaccent 19 \theta}
\def\lam'{\mathaccent 19 \lambda}
\def\varet'{\mathaccent 19 \vartheta}
\def\rh'{\mathaccent 19 \rho}
\def\ph'{\mathaccent 19 \phi}
\def\xb'{\mathaccent 19 {\bar{x}}}

\def\l{{\lambda}}

\def\be{\begin{equation}}
\def\ee{\end{equation}}

\newcommand{\bea}{\begin{eqnarray}}
\newcommand{\eea}{\end{eqnarray}}

\def\cneg{ \chi_{_{\rm neg}}}
\def\cmin{ \chi_{_{\rm min}}}
\def\cmax{ \chi_{_{\rm max}}}

\def\om{\omega}
\def\de{\delta}
\def\De{\Delta}
\def\r {\rho}
\def\a {\alpha}
\def\b {\beta}
\def\s {\sigma}

\def\g {\gamma}
\def\p{\phi}
\def\vp {\varphi}

\def\la{\label}
\def\e{\epsilon}
\def\ov{\over}
\def\tr{{\rm tr}}

\def\hg {{\hat \gamma}}
\def\cJ{{\cal J}}

\newcommand{\alg}[1]{\mathfrak{#1}}
\newcommand{\su}{\alg{su}}

\newcommand{\AdS}{{\rm  AdS}_5\times {\rm S}^5}

\author{
Dmitri Bykov$^{a,b}$\footnote{Email: dbykov, frolovs@maths.tcd.ie}
 \  and Sergey Frolov$^{a}$\footnote{Correspondent fellow at Steklov Mathematical Institute,
Moscow.}
 \\ $^a$ {\it School of Mathematics,
Trinity College, Dublin 2, Ireland}\\$^{b}$ {\it Steklov
Mathematical Institute, Moscow, Russia} }

\abstract{We consider giant magnons propagating in a
 $\g$-deformed $\AdS$ background obtained from $\AdS$ by means of a chain of TsT transformations.  We point out that in the light-cone gauge and in the infinite $J$ limit the deformed and undeformed string models share the same magnon dispersion relation,  the $\su(2|2)\oplus \su(2|2)$-invariant world-sheet S-matrix and the dressing factor. The $\g$-dependence in the limit is only due to different level-matching conditions.
We consider the reduction of the deformed model to $R\times S^3$ and determine the leading $\g$-dependence of the dispersion relation for a finite $J$ giant magnon.

}

\title{Giant magnons in TsT-transformed $\AdS$}

\preprint{
          \smaller{\smaller{\smaller{TCDMATH 08-06}}}}

\begin{document}

%\newpage

\renewcommand{\thefootnote}{\arabic{footnote}}
\setcounter{footnote}{0}
\section{Introduction and summary}
An interesting example of the AdS/CFT duality \cite{M}  between gauge and string theory models with reduced supersymmetry is provided by an exactly marginal deformation of ${\cal N}=4$ super Yang-Mills theory \cite{LEST} and string theory on a deformed $\AdS$ background suggested in \cite{LM}. The deformed models depend on a continuous complex parameter $\b$, and are often called $\b$-deformed. If $\b\equiv\g$ is real the deformed string background can be derived from $\AdS$ by
using a TsT transformation which is a combination of
a T-duality  on one angle variable,
a shift of another isometry variable, followed by the second T-duality on the first angle \cite{LM,F}. Moreover, since $S^5$ has three isometry directions, a chain of TsT transformations can be used to  construct a regular
three-parameter deformation
of $\AdS$ dual to a non-supersymmetric deformation of ${\cal N}=4$ SYM \cite{F}.
The  Lagrangian of the  $\g_i$-deformed gauge theory can be obtained
from the undeformed one by replacing the usual product by the associative $*$-product
 \cite{LM, F, BR}. The resulting model is conformal in the planar limit to any order of  perturbation theory \cite{Ko}.

 \smallskip

Another important property of a TsT transformation is that it preserves the classical  integrability of string theory on $\AdS$ \cite{F}. In particular the Lax  pair for strings on $\AdS$ \cite{BPR}
and a TsT transformation can be used to find a Lax pair for strings on a deformed background \cite{F,AAF}. Moreover, the Green-Schwarz action for strings on $\AdS$ is mapped under a TsT transformation to a string action on the $\g$-deformed background providing a nontrivial example of non-supersymmetric Green-Schwarz action for strings on RR backgrounds \cite{AAF}. In fact in the Hamiltonian (first-order) formalism the Green-Schwarz action for strings on  the $\g$-deformed background is canonically equivalent to the action for strings on $\AdS$ satisfying  quasi-periodic or twisted boundary conditions \cite{F,AAF}. The twists however are quite unusual because they depend on charges carried by a string and are given by linear combinations of products of the deformation parameters and $\su(4)$ charges.

 \smallskip

This also implies that in the light-cone gauges of \cite{AF04,FPZ}  the string dynamics  on both the $\g$-deformed background and $\AdS$ is described by the same Hamiltonian density. The $\g$-dependence enters only through the twisted boundary conditions and the level-matching condition which is modified because a closed string in the deformed background in general corresponds to an open string in $\AdS$.
Correspondingly, in the decompactification limit where one of the $\su(4)$ charges, say $J$, is sent to infinity while the string tension and the deformation parameters are kept fixed the dependence of the light-cone Hamiltonian on the deformation parameters disappears  because in this limit all physical fields must vanish at the space infinity\footnote{A $\g$-dependence remains in the pp-wave \cite{BMN} and spinning string \cite{FT} limits because in these limits the effective length $J/\sqrt\l$ and the twists $\sim \g_i J_k$ are kept fixed, and therefore the string sigma model is defined on a circle with fields obeying quasi-periodic boundary conditions. The pp-wave limits of the deformed backgrounds were discussed in \cite{Niarchos,Koch,Mateos}, and  the finite-gap integral equations \cite{KMMZ} describing spinning strings in the $\g$-deformed $\su(2)$ sector were derived in \cite{FRT1}.}. As a result, if one considers the light-cone gauge-fixed string sigma model off-shell, that is if one does not impose the level-matching condition then the deformed string model is indistinguishable from the undeformed one, and they share the same magnon dispersion relation \cite{BDS},  the $\su(2|2)\oplus \su(2|2)$-invariant world-sheet S-matrix \cite{S,B,AFZZ} and the dressing factor \cite{AFS}-\cite{BES}. Therefore,  the $\g$-dependence in the decompactification limit is only due to the level-matching condition.

 \smallskip

Thus, to see the dependence of the off-shell spectrum of the model on the deformation parameters one should analyze it for finite values of the $\su(4)$ charges. The leading dependence can then be captured by the asymptotic Bethe ansatz which would
differ from the usual one \cite{BS} only by the twists reflecting the non-periodic boundary conditions for finite $J$. This conclusion is also confirmed by the one-loop considerations in the $\g$-deformed gauge theory \cite{RR,BECH,BR} where it is shown that
the one-loop integrability of ${\cal N}=4$ SYM \cite{MZ} is preserved by the deformation, and the corresponding one-loop Bethe ansatz involves the same twists that appear in string theory \cite{BR}.  In the asymptotic approximation the dispersion relation is not modified and  the twists lead to a very mild modification of the string spectrum which basically reduces to $\g$-dependent  shifts of string mode numbers, see \cite{LM,FRT1,FRT2} for some examples.

 \smallskip

The asymptotic Bethe ansatz is not exact and for finite $J$ one expects to find a non-trivial $\g$-dependence already in the large string tension limit where classical string considerations can be used. In particular, it is interesting to determine how the dispersion relation for a giant magnon \cite{HM} depends on the deformation parameters.
In the infinite $J$ limit a giant magnon is dual to a gauge theory spin chain magnon, and
 in the conformal gauge it can be identified with an open string solution
of the sigma model reduced to $R\times S^2$. The end-points of the open string move along the equator of $S^2$ parametrized by an angle $\p$, and
 the momentum $p$ carried by the dual spin chain magnon is equal to the difference in the angle $\p$ between the two end-points of the string  \cite{HM}.  On the other hand
  in a light-cone gauge a giant magnon is identified with a world-sheet soliton
and the momentum $p$ is equal to the world-sheet momentum $p_{\rm ws}$ of the soliton  \cite{AFZ}. For finite $J$ the equality between $p$ and $p_{\rm ws}$ holds only in the light-cone gauge $t=\tau\,,\ p_\p=1$  \cite{AFZ}.

 \smallskip

In this paper we  determine the leading $\g$-dependence of the dispersion relation for a finite $J$ giant magnon. We use the conformal gauge and the string sigma model reduced to $R\times S^3$ which in the deformed case is the smallest consistent reduction due to the twisted boundary conditions. Even for the three-parameter deformation the reduced model depends only on one of the parameters which we denote $\g$. Since there are two isometry angles $\p_1$ and $\p_2$ a solution of the reduced model can have two non-vanishing charges $J_1$ and $J_2$. A giant magnon  is then an open string solution of the model which carries only one charge $J\equiv J_1$.
The momentum $p$ of the magnon is correspondingly identified with the difference in the angle $\p_1$ between the two end-points of the open string because in the light-cone gauge $t=\tau\,,\ p_{\p_1}=1$ it is equal to the world-sheet momentum of a soliton.  The second angle $\p_2$ satisfies a twisted boundary condition which can be found by using the general formulas from \cite{F}
 \bea\nonumber
\De\p_2=2\pi (n_2
-\g J )\,,\quad n_2\in{\mathbf Z}\,,\eea
where $n_2$ is an integer winding number of the string in the second isometry direction of the deformed sphere $S^3_\g$. Collecting all the requirements together, we conclude that
a $\g$-deformed giant magnon can be identified with an open string in $R\times S^3$ satisfying the following conditions
 \bea \nonumber
\De\p_1=p\, ,\quad \De\p_2=2\pi (n_2
-\g J )\, ,\quad J_1=J\,,\quad J_2=0\,.\eea
We analyze the equations of motion and find that a solution exists only for one integer $n_2$ which obeys  the condition
$
|n_2 - \g J| \le {1\ov 2}\,,
$
and therefore there is only one deformation of a giant magnon solution in $R\times S^2$.
Then,
the leading correction to the dispersion relation in the large $J$ limit has the following form
\begin{equation}\nonumber
E-J=2g \sin{\frac{p}{2}}
\left(1-\frac{4}{e^{2}}\, \sin^{2}\frac{p}{2} \,\cos{\Phi}\,
e^{-\frac{\mathcal{J}}{\sin p/2}}+...\right)\,,\quad \Phi= \frac{2\pi(n_2-\g J)}{2^{3/2} \cos^{3}\frac{p}{4}}\,,
\end{equation}
where
$g={\sqrt\l\ov 2\pi}$ is the string tension, and $\cJ =J/g$.
The  formula reduces  in the limit $\g\to 0$ (or $\Phi \to 0$) to the one obtained in \cite{AFZ}.  In the large $J$ limit the $\g$-dependence disappears in agreement with the discussion above, and if $\g$ is kept fixed then the winding number $n_2$ goes to infinity too.

The deformed theory has less supersymmetry, and one expects that the energy of a $\g$-deformed magnon would be higher than the energy of the undeformed one with the same momentum and charge. It is indeed the case because $\cos\Phi < 1$.

 It would be interesting to understand how to reproduce the dispersion relation  by using L\"uscher's approach \cite{Luscher}. This would generalize the computation performed in \cite{Janik} to the deformed case.
The dispersion relation has a peculiar $\g$-dependence for finite $J$, and it is not quite clear how such a dependence follows from the S-matrix approach. This would require to generalize L\"uscher's formulas to the case of the nontrivial twisted boundary conditions.

Our consideration can be generalized to solutions carrying several spins, see
\cite{Ryo,Min,Klose} for recent discussions of the undeformed model. It would be also interesting to compute the one-loop quantum correction generalizing the considerations in \cite{Gromov,Heller}.

 \smallskip

In section 2 we discuss possible giant magnon solutions in the deformed background and explain how they can be mapped to open strings in $\AdS$. In section 3 we sketch the derivation of the leading correction to the dispersion relation in the large $J$ limit and discuss its structure. The details of the derivation can be found in Appendix.

%%%%%%%%%%%%%%%%%%%%%%%%%%%%%%%%%%%%%%%%%%%%%%%%%%%%%%%%%%%%%%%%%%%%%%%%%%%%%%%%%
\section{The $\gamma$-deformed giant magnon}
%%%%%%%%%%%%%%%%%%%%%%%%%%%%%%%%%%%%%%%%%%%%%%%%%%%%%%%%%%%%%%%%%%%%%%%%%%%%%%
The bosonic part of the Green-Schwarz action for strings on the $\g$-deformed $\AdS$ background \cite{AAF} reduced to $R\times S^5_\g$ can be written in the following form
\bea\nonumber
S = -{g\over 2}\int_{-r}^r  d\s d\tau\Big[ \g^{\a\b}\left(-\partial_\a t\partial_\b t +\partial_\a \r_i\partial_\b \r_i + G\,
\r_i^2\partial_\a\vp_i\partial_\b\vp_i+ G\,
\r_1^2\r_2^2\r_3^2\big(\hg_i\partial_\a
\vp_i\big)\big(\hg_j\partial_\b\vp_j\big)\right) &&\\
\la{a1} - 2\, G\, \e^{\a\b}\left(\hg_3\r_1^2\r_2^2\partial_\a
\vp_1\partial_\b\vp_2 +\hg_1 \r_2^2\r_3^2\partial_\a\vp_2\partial_\b
\vp_3 +\hg_2 \r_3^2\r_1^2\partial_\a\vp_3\partial_\b\vp_1 \right)\Big]\, . \ \ \ \ \ \  &&
\eea
Here  $g ={R^2\ov \a'}= {\sqrt\l\ov 2\pi}$ is the string tension, and  $\g^{\a\b}= \sqrt{-h}h^{\a\b}$ where
$h^{\a\b}$ is a world-sheet metric with Minkowski signature. The function $G$ is defined as follows
\bea
G^{-1} = 1 +  \hg_3^2  \rho_1^2 \rho_2^2 +  \hg_1^2
\rho_2^2 \rho_3^2 +  \hg_2^2  \rho_1^2 \rho_3^2
~,\quad \ \ \ \ \sum_{i=1}^3\rho_i^2=1\,,
\eea
and $\vp_i$ are the three isometry  angles of the deformed $S^5_\g$. The deformation parameters $\hg_i$ are kept fixed in the string sigma model perturbation theory, and are related to the parameters $\g_i$ which appear in the dual gauge theory as $\hg_i =  2\pi g\g_i = \sqrt\l \g_i$. The standard $\AdS$ background is recovered  after setting the deformation parameters $\hg_i$  to zero. For equal $\hg_i=\hg$ this becomes the supersymmetric background of \cite{LM}, and the deformation
parameter $\g$ enters the ${\cal N}=1$ SYM superpotential as follows
$W= h\, \tr( e^{ i  \pi \g} \Phi_1\Phi_2 \Phi_3 - e^{- i  \pi \g}
 \Phi_1\Phi_3 \Phi_2  )$.

\medskip

The TsT transformations
that map  the $AdS_5\times S^5$ string theory to the
$\g_i$-deformed string theory allow one to relate
 the angle variables $\p_i$ of $S^5$
 to  the angle variables $\vp_i$ of the $\g$-deformed geometry.
The relations take their simplest form being expressed in terms of
the momenta $p_i, \pi_i$ conjugate to $\p_i, \vp_i$,
respectively\footnote{Here we use definitions of momenta $p_{i}$,
which differ by a factor of $2\pi$ from those of \cite{F}, therefore
we have an extra $2\pi$ in (\ref{rel3}).} \cite{F} \bea p_i &=&
\pi_i \, ,
\label{rel2}\\
\r_i^2\,\p_i' &=& \r_i^2 ( \vp_i' - 2\pi\epsilon_{ijk}\g_j p_k )\ ,
\ \ \ \ \  i=1,2,3 \ , \label{rel3} \eea where in (\ref{rel3})  we
sum only in $j,k$. The relation (\ref{rel2}) implies that the $U(1)$
charges $J_i=\int d\s p_i$ are invariant under a TsT transformation.

Assuming that none of the ``radii'' $\r_i$ vanish on a string
solution, we  get \bea \p_i' = \vp_i' - 2\pi\epsilon_{ijk}\g_j p_k\
. \label{rel4} \eea Integrating eq.(\ref{rel4}) and taking into
account that \bea\la{mn} \De\vp_i=\vp_i(r)-\vp_i(-r)=2\pi n_i\, ,
n_i\in \mathbb{Z} \eea for a closed string in the $\g$-deformed
background, we obtain the twisted boundary conditions for the angle
variables $\p_i$ of the original $S^5$ space \bea \label{tbcb}
\De\p_i=\phi_i(r)-\phi_i(-r)=2\pi (n_i -\nu_i )\, ,\quad \nu_i =
\e_{ijk}\gamma_jJ_k\,,\quad  J_i = \int^{r}_{-r}  d \s \  p_i  .\eea

It is clear that if the twists $\nu_i$
 are not integer then a closed string in the deformed
 geometry is mapped to an open string in $\AdS$. A giant magnon solution in this respect does not differ essentially from a closed string in AdS$_5\times$S$^5_\g$.
 It corresponds to an open string in the deformed geometry, and its image in $\AdS$ is an open string too. The only difference is that not all of the winding numbers $n_i$ are integer for a giant magnon solution. In fact one linear combination of the winding numbers should be identified with the momentum $p$ carried by the giant magnon.

 \smallskip

 To determine the linear combination we notice that in the infinite $J\equiv J_1+J_2+J_3$ limit the end-points of a giant magnon should move with the speed of light along a null geodesic of the background  \cite{HM}.  In the undeformed case any geodesics is just a big circle of $S^5$, and the solution is described by a soliton of the string sigma model reduced to $R\times S^2$. The momentum carried by the soliton is identified with the difference in the angle $\p$ between the two end-points of the string where $\p$
 parametrizes the equator of $S^2$  \cite{HM}.  In the light cone gauge $t=\tau\,,\ p_\p=1$ the momentum $p$ is equal to the world-sheet momentum of the giant magnon solution and because of that the identification can be also used for finite $J$ \cite{AFZ}.

 \smallskip

 In the $\g$-deformed background there are infinitely many inequivalent geodesics which correspond to solutions of the  Neumann-Rosochatius
 integrable system \cite{FRT2} (which also describes multi-spin string solutions \cite{AFRT,ART}), and
 one should choose only those which give the minimum energy satisfying the BPS condition $E=J$. These geodesics were described in \cite{FRT2} where it was shown that for generic values of $\g_i$ there are three BPS states which have only one of the three charges $J_i$ nonvanishing. Choosing for definiteness  the nonvanishing charge to be $J_1=J$, the BPS state corresponds to the geodesics parametrized by the angle $\vp_1$ and having $\r_1=1\,,\ \r_2=\r_3=0$. An infinite $J$ giant magnon with the end-points moving along the geodesics is then a solution of the string sigma model reduced to $R\times S^3_\g$ where $S^3_\g$ is obtained from the deformed $S^5_\g$ by setting $\r_3=0$. The momentum $p$ carried by the soliton is identified with the difference
 $\De\vp_1=\vp_1(r)-\vp_1(-r)$. In fact it is easy to see that the TsT transformation maps  the infinite $J$ giant magnon solution of the undeformed model to
 the $\g$-deformed giant magnon, and therefore the infinite $J$ dispersion relation is not modified, and has no $\g$ dependence. For finite $J$ however the dispersion relation gets a nontrivial $\g$-dependence which we determine in the next section. This follows from the fact that for the magnon solution $J_2=J_3=0$, and therefore the twist $\nu_1=0$, and the corresponding angles $\p_1$ and $\p_2$ of the undeformed $S^3$ satisfy the following twisted boundary conditions
 \bea \label{tbcb2}
\De\p_1=\phi_1(r)-\phi_1(-r)=p\, ,\quad \De\p_2=\phi_2(r)-\phi_2(-r)=2\pi (n_2
-\g J )\, ,\eea
where $\g \equiv \g_3\,,\ J\equiv J_1$. As a result the dispersion relation for the finite $J$ $\g$-deformed giant magnon depends on $p, J$ and $\de \equiv 2\pi(n_2-\g J )$.
 To find the dispersion relation one can either use the conformal gauge \cite{HM} or the light-cone gauge \cite{AFZ}.

\smallskip

Let us also mention that in the case where the deformation parameters satisfy the relations $\g_i = c\, k_i$ where $c$ is any real number and  $k_i$  are arbitrary integers,  there is another family of BPS states with the following charges \cite{FRT2}
\bea
J_i = k\, k_i\sim \g_i\, ,
\la{rel8}
\eea
where (in quantum theory) $k$ is any integer. In particular, in the supersymmetric case $\g_i=\g$ the BPS states are the states $(J/3,J/3,J/3)$ with three equal charges. Since $J_i\sim\g_i$ for these BPS states the twists $\nu_i$ vanish and both the $\g$-deformed giant magnon and its TsT image satisfy the same twisted boundary conditions which take the simplest form in terms of the following new angle variables and their conjugate momenta
\bea
\psi_1 = k_1 \p_1 +  k_2 \p_2 +  k_3 \p_3\,, &&\quad \pi_1 = {p_1+p_2+p_3\ov k_1+k_2+k_3}\,,\\
 \psi_2 = k_1 \p_1 -  (k_1+k_3) \p_2 +  k_3 \p_3\,,&&\quad \pi_2 = {k_2 p_1-k_1p_2\ov k_1(k_1+k_2+k_3)}\,,\\
 \psi_3 = k_1 \p_1 +  k_2 \p_2 -  (k_1+k_2) \p_3\,, &&\quad \pi_3 = {k_3 p_1-k_1p_3\ov k_1(k_1+k_2+k_3)}\,.
\eea
Then, the giant magnon solution with the charges satisfying (\ref{rel8}) satisfies the following boundary conditions
\bea
\De\psi_1 = p\,,\quad \De\psi_2 = 0\,,\quad \De\psi_3 = 0\,.
\eea
Since the boundary conditions do not depend on $\g_i$ in the classical theory the dispersion relation for the giant magnon does not depend on the deformation parameters either. A disadvantage of this giant magnon solution is that the corresponding Bethe ansatz is not known.

\section{Finite $J$ dispersion relation}

To determine the dispersion relation we impose the conformal gauge $\g^{\a\b} = \mbox{diag}(-1,1)$,  set $t=\tau$,  and use the following parametrization of $S^3$
\bea
x_i^2=1\,,\ x_1 + i x_2 = \r_1 e^{i\p_1}\,,\ x_3 + i x_4 = \r_2 e^{i\p_2}\,,\ \r_2^2 = 1-\r_1^2=\chi\,.~~~
\eea
Then the sigma model action for strings on $R\times S^3$ takes the following form
\bea\nonumber
S = -{g\over 2}\int_{-r}^r  d\s d\tau\left({\partial_\a \chi\partial^\a \chi\ov 4\chi(1-\chi)} +
(1-\chi)\partial_\a\p_1\partial^\a\p_1+  \chi\partial_\a\p_2\partial^\a\p_2  \right)\, . \ \ \ \ \ \
\eea
and  solutions of the equations of motion should also  satisfy the
Virasoro constraints
\bea\la{Vcs1} &&{\dot{\chi}^2 + \chi'^2\ov 4\chi(1-\chi)}
+(1-\chi)\left(
\dot{\p}_1^2 + \p_1'^2\right) +\chi\left( \dot{\p}_2^2 + \p_2'^2\right) = 1\,,\\\la{Vcs2}
&&{\dot{\chi}\chi'\ov 4\chi(1-\chi)} +(1-\chi)\dot{\p}_1\p_1'+ \chi\dot{\p}_2\p_2'= 0\, .
\eea
Since $t=\tau$ the range of $\s$ is related to the space-time energy $E$ of a solution as follows
\bea
2r ={E\ov g}\equiv  {\cal E}\,.
\eea
The
two charges $J_1\equiv J$ and $J_2$ corresponding to shifts of
$\p_1$ and $\p_2$ are
\bea\la{Jcg1} J =g\int_{-r}^{ r}\, {\rm d}\s\, (1-\chi)\, \dot{\p}_1 \,,\quad J_2
= g\int_{-r}^{ r}\, {\rm d}\s\, \chi\,
\dot{\p}_2\,.\eea
As was discussed in the previous section, the $\g$-deformed giant magnon solution
has only one nonvanishing charge $J$, and
the angles $\p_1$ and $\p_2$ satisfy the following twisted boundary conditions
 \bea \label{tbcb3}
\De\p_1=\phi_1(r)-\phi_1(-r)=p\, ,\quad \De\p_2=\phi_2(r)-\phi_2(-r)=\de\, ,\eea
where $\de = 2\pi(n_2-\g J) $,  $\g= \g_3$ and $n_2$ is the winding number in the $\vp_2$ direction of the deformed $S^5_\g$ . It is worth mentioning that the dependence on $\g$ and $n_2$ comes only through their linear combination $\de$ which in fact plays the role of the deformation parameter.

The problem of finding a finite $J$ giant magnon solution is thus basically equivalent to the problem of finding a two-spin giant magnon solution discussed in appendix C of \cite{AFZ}, and can be solved by using a similar ansatz
\bea
\p_1(\s,\tau) &=& \om\tau + {p\ov 2r}(\s - v\tau) + \p(\s - v\tau)\,,\\
\p_2(\s,\tau) &=& \nu\tau + {\de\ov 2r}(\s - v\tau)+ \a(\s - v\tau)\,,\\
\chi(\s,\tau) &=& \chi(\s - v\tau)\,,
\eea
where $\chi(\s)$, $\p(\s)$ and $\a(\s)$ satisfy the periodic boundary conditions.

Substituting the ansatz into the equations of motion, integrating the equations for $\p$ and $\a$ once, and using the Virasoro constraint (\ref{Vcs1}) , we get the following three equations
\begin{eqnarray}
  \phi' &=&f_0+\frac{f_1}{1-\chi}\,, \quad
  \alpha' = a_0+\frac{a_1}{\chi}\,, \\
  \la{eom1}
\kappa^2\,\chi'^{2}&=& (\chi -\cneg)(\chi-\cmin)(\cmax-\chi )\,,
\end{eqnarray}
where the constants in the equations are functions of $\om,\nu,v,p,\de$, and
$\cneg,\cmin,\cmax$ are ordered as
$\cneg\leq 0\leq \cmin < \cmax$. Moreover,  giant magnon solutions exist only if $\cmax\le 1$ and for these solutions $\cmin\le\chi\le\cmax$, see Appendix for detail.

If the deformation parameter $\de$ goes to 0 then $\cneg\,,\, a_0\,,\,a_1$
approach 0 too, and we recover the equations of motion for a finite $J$ undeformed giant magnon \cite{AFZ}.

For any value of $\de$ we can always choose the initial conditions so that $\chi(\s)$ is an even function and  $\p(\s)$ and $\a(\s)$ are odd functions of $\s$, and since they are also periodic functions, we can always look for a solution satisfying the following boundary conditions
\bea\la{bcon}
&&\chi(-r)=\chi(r)=\cmin\,,\quad \chi(0)=\cmax\,,\quad \chi(-\s)=\chi(\s)\,,\\\nonumber
&& \p(-r)= \p(0)=\a(-r)= \a(0)=0\,,\quad  \p(-\s)=-\p(\s)\,,\quad  \a(-\s)=-\a(\s)\,.~~~~~
\eea
Due to the conditions we can restrict our attention to the half of the string from $-r$ to $0$, and since $\chi$ is an increasing function on this interval we can also
replace integrals over $\s$ by integrals over $\chi$ from $\cmin$ to $\cmax$.
Then a solution is completely determined by the following five equations which are analyzed in detail in Appendix
\begin{eqnarray}
\nonumber &&\text{Periodicity of }\phi:\;\;\; r\, f_0+f_1\int\limits_{\cmin}^{\cmax}\frac{d\chi}{(1-\chi)|\chi'|}=0\,, \\
\nonumber &&\text{Periodicity of }\alpha:\;\;\; r\, a_0+a_1\int\limits_{\cmin}^{\cmax}\frac{d\chi}{\chi|\chi'|}=0\,, \\
\nonumber &&\text{Charge } \cJ\equiv {J_{1}\ov g}:\;\;\; \cJ =-2r\,v\, f_1 +
\frac{\om}{1-v^{2}}\int\limits_{\cmin}^{\cmax} d\chi\,\frac{1-\chi}{|\chi'|}\,, \\
\nonumber &&\text{Charge }J_{2}=0:\;\;\; 0 =
-2r\,v\,a_1+{\nu\ov 1-v^2}\int\limits_{\cmin}^{\cmax}
  d\chi\,\frac{\chi}{|\chi'|}\,,\\
\nonumber
&&\text{Length of string:}\;\;\;
\int_{-r}^0 d\s =r=\int\limits_{\cmin}^{\cmax}\,\frac{d\chi}{|\chi'|}\,,
\end{eqnarray}
where all constants should be expressed in terms of the charge $\cJ$, the soliton momentum $p$ and the deformation parameter $\de$.

\medskip

The dispersion relation can be found in the large $\cJ$ limit as an expansion in $e^{-\frac{\mathcal{J}}{\sin(p/2)}}$, and up to the first correction it has the following form ($0\le p \le\pi$)
\begin{equation}\label{mainform}
E-J=2g \sin{\frac{p}{2}} \left(1-\frac{4}{e^{2}}\,
\sin^{2}\frac{p}{2} \,\cos{\Phi}\,
e^{-\frac{\mathcal{J}}{\sin p/2}}+...\right)\,,
\end{equation}
where
\begin{equation}\label{Ph}
\Phi=\frac{\de}{2^{3/2} \cos^{3}\frac{p}{4}} = \frac{2\pi(n_2-\g J)}{2^{3/2} \cos^{3}\frac{p}{4}}\,.
\end{equation}
The  dispersion relation in the $\g$-deformed model reduces  in the limit $\de\to 0$ (or $\Phi \to 0$) to the one obtained in \cite{AFZ}.

Some remarks are in order.
\begin{enumerate}
\item We see that in the limit $\cJ\to\infty$ the dispersion relation is independent of the deformation parameter. This is contrary to papers \cite{Kh,Bobev} where it was claimed that the momentum is shifted by the deformation parameter $2\pi \g$. As was discussed in the previous section, $2\pi\g$ is identified with $\hg/g$, and therefore the shift by $\g$ cannot be seen in classical theory in any case. It would be a one-loop effect, and the discussion in the Introduction indicates that the momentum $p$ is not shifted at one loop at all but one should take into account that in quantum theory magnons carry other charges of order one, and therefore $p=\De\p_1$ is not equal to $p_{\rm ws}=\De\vp_1$. According to (\ref{tbcb}), if we have several (or just one) magnons with the total charges $J_2, J_3$ then the momenta are related as $p=p_{\rm ws}+ 2\pi\g_3 J_2 - 2\pi\g_2 J_3 $. If the state is physical then the total world-sheet momentum $p_{\rm ws}$ should vanish leading to the condition $p= 2\pi\g_3 J_2 - 2\pi\g_2 J_3$ (up to an integer multiple of $2\pi$). This condition is equivalent to the cyclicity constraint in the twisted Bethe ansatz \cite{BR}.

\item Since $\cos\Phi < 1$ the energy of a $\g$-deformed magnon is higher than the energy of the undeformed one with the same momentum and charge. That is what one should expect because the deformed theory has less supersymmetry.

\item The derivation of the dispersion relation performed in Appendix shows that a giant magnon solution exists if $\Phi$ satisfies the restriction
\bea\la{rest1} -\pi\le \Phi \le \pi\,, \eea and therefore if we require
a solution to exist for all values of $p$ from $-\pi$ to $\pi$ the
parameter $\de$ must also satisfy the same restriction
\bea\la{rest2} -\pi\le \de \le \pi\ \ \Longleftrightarrow\ \ |n_2 - \g
J| \le {1\ov 2}\, . \eea This means that $n_2$ is the integer closest
to $\g J$.  We see that for any $\g J$ there is only one integer
$n_2$ which satisfies the condition, and therefore there is only one
deformation of a giant magnon solution in $R\times S^2$. If the
fractional part of $\g J$ is less than $1/2$ then $n_2$ is equal to
the integer part of $\g J$, and if the fractional part of $\g J$ is
greater than $1/2$ then $n_2$ is equal to the integer part of $\g J
+1$.

\item For small enough values of $p$ however the first-order perturbation theory in $e^{-\frac{\mathcal{J}}{\sin(p/2)}}$ allows one to have two or three integers satisfying the restriction (\ref{rest1}): $n_2$ satisfying (\ref{rest2}), and  $n_2\pm 1$. We expect that the latter possibilities will be ruled out at higher orders of the perturbation
theory. Anyway, according to (\ref{mainform}) their energies would
be higher than the energy of the main solution.

\end{enumerate}

%%%%%%%%%%%%%%%%%%%%%%%%

\section*{Acknowledgements}
We thank  Gleb Arutyunov  for  discussions. The work of D.B. was
supported by the EU-RTN network {\it Constituents, Fundamental
Forces and Symmetries of the Universe} (MRTN-CT-2004-512194), in
part by grant of RFBR ¹ 08-01-00281-a and in part by grant for the
Support of Leading Scientific Schools of Russia NSh-795.2008.1.
 The work of S.F. was supported in part by the Science
Foundation Ireland under Grant No. 07/RFP/PHYF104.

%%%%%%%%%%%%%%%%%%%%%%%%%%%%%%%%%%%%%

\appendix

%%%%%%%%%%%%%%%%%%%%%%%%%%%%%%%%%%%%%%%%%%%%

\section{The motion on $\gamma$-deformed $S^{3}$.}

The metric of $AdS_{5}\times S^{5}$, reduced to the
$\mathbb{R}\times S^{3}$ takes the following form:
\begin{equation}\label{1}
ds^{2}=-dt^{2}+\frac{d\chi^{2}}{4\chi(1-\chi)}+(1-\chi)d\phi_{1}^{2}+\chi
d\phi_{2}^{2}.
\end{equation}
We will be looking for a solution of the equations of motion in the
following form:
\begin{eqnarray}
  &&\phi_{1}(\sigma, \tau) = \omega\tau + \frac{p}{2r}(\sigma-v \tau)+\phi(\sigma-v\tau); \\
  &&\phi_{2}(\sigma,\tau) = \nu\tau
  +\frac{\delta}{2r}(\sigma-v\tau)+\alpha(\sigma-v\tau);\\&&
  \chi(\sigma,\tau)=\chi(\sigma-v\tau),
\end{eqnarray}
where $\delta = 2\pi(n_{2}-\gamma
  J_{1})$ and $\phi(\sigma-v\tau)$, $\alpha(\sigma-v\tau), \chi(\sigma-v\tau)$ satisfy periodic boundary conditions.

Substituting the ansatz into the equations of motion, integrating
the equations for $\p$ and $\a$ once, and using the Virasoro
constraints (\ref{Vcs1}) , we get the following equations:
\begin{eqnarray}\label{phieq}
  &&\phi' = -(\frac{v\omega}{1-v^2}+\frac{p}{2r})-\frac{vA_{1}}{1-v^2}\frac{1}{1-\chi}
  \\ \label{aleq}
  &&\alpha' = -(\frac{v\nu}{1-v^2}+\frac{\delta}{2r})-\frac{vA_{2}}{1-v^2}\frac{1}{\chi}
  \\ \label{chieq}
  &&\frac{(1-v^2)^2}{4}\chi'^{2}=
\kappa_{0}+\kappa_{1}\chi+\kappa_{2}\chi^{2}+\kappa_{3}\chi^{3}\\
&&\omega A_{1}+ \nu A_{2}+ 1 =0.
\end{eqnarray}
The constants $\kappa_{i}$ are as follows:
\begin{eqnarray}
  \kappa_{0} &=& -v^{2}\,A_{2}^{2} \\
  \kappa_{1} &=& 1-\omega^{2}+v^{2}(1+A_{2}^{2}-A_{1}^{2}) \\
  \kappa_{2} &=& -1-\nu^{2}+2\omega^{2}-v^{2} \\
  \kappa_{3} &=& \nu^{2}-\omega^{2},
\end{eqnarray}
Thus, in the notation of section 3 one may write
\begin{eqnarray}
\nonumber
 && f_{0} = -(\frac{v\omega}{1-v^2}+\frac{p}{2r});\;\;f_{1}=-\frac{vA_{1}}{1-v^2}; \\ \nonumber
&&
a_{0}=-(\frac{v\nu}{1-v^2}+\frac{\delta}{2r});\;\;a_{1}=-\frac{vA_{2}}{1-v^2};\\
\nonumber &&\kappa = \frac{1-v^2}{2 \sqrt{\omega^2-\nu^2}}.
\end{eqnarray}
We also have the following expressions for the charges\footnote{From
these expressions one can derive a linear relation between
$E,\mathcal{J},\mathcal{J}_{2}$:
$$\frac{1-v^2}{\mathcal{E}}\left(\frac{\mathcal{J}}{\omega}+\frac{\mathcal{J}_{2}}{\nu}\right)=1+v^{2} \left(\frac{A_{1}}{\omega}+\frac{A_{2}}{\nu} \right).$$}:
\begin{eqnarray}
  \mathcal{J} &=& \frac{1}{1-v^2} (2rv^{2}A_{1}+\omega\int\limits_{-r}^{r} d\sigma\,(1-\chi)) \\
  \mathcal{J}_{2} &=& \frac{1}{1-v^2}
  (2rv^{2}A_{2}+\nu\int\limits_{-r}^{r}
  d\sigma\,\chi)=0.
\end{eqnarray}
Our equations can be written in the following form:
\begin{eqnarray}
\label{int1}  &&\text{Periodicity of }\phi:\;\;\;  \frac{rv\omega}{1-v^{2}}+\frac{p}{2}=-\frac{v\,A_{1}}{1-v^2}\int\limits_{\cmin}^{\cmax}\frac{d\chi}{(1-\chi)|\chi'|}; \\
\label{int2}  &&\text{Periodicity of }\alpha:\;\;\;
\frac{rv\nu}{1-v^{2}}+\pi \delta
=-\frac{v\,A_{2}}{1-v^2}\int\limits_{\cmin}^{\cmax}\frac{d\chi}{\chi|\chi'|};\\
\label{int3}  &&\text{Charge }\cJ\equiv {J_{1}\ov g}:\;\;\; \mathcal{J} = \frac{2}{1-v^{2}} \left(rA_{1}v^{2}+\omega\int\limits_{\cmin}^{\cmax} d\chi\,\frac{(1-\chi)}{|\chi'|})\right); \\
\label{int4}  &&\text{Charge }\mathcal{J}_{2}\equiv {J_{2}\ov
g}=0:\;\;\; 0 = rv^{2}A_{2}+\nu\int\limits_{\cmin}^{\cmax}
  d\chi\,\frac{\chi}{|\chi'|},
\end{eqnarray}
and the periodicity condition for $\chi$ which in this case takes
the form
\begin{equation}\label{int5}
\text{Length of string:}\;\;\; \int_{-r}^0 d\s
=r=\int\limits_{\cmin}^{\cmax}\,\frac{d\chi}{|\chi'|}\;.
\end{equation}
We have called the real roots of the equation $\cneg,\cmin,\cmax$
with the following ordering $\cneg\leq 0\leq \cmin < \cmax$.
Moreover, for the consistency of our approach we have to require
that $\cmin, \cmax \in[0,1)$, which will be justified by the
solution. The fact that in the large $J$ expansion one of the roots
is negative can be easily proven. Indeed, in the strict $J\to
\infty$ limit it follows from the work \cite{AFZ} that $\omega=1, \;
\nu=0$, therefore the leading coefficient $\kappa_{3}$ of the
polynomial in the r.h.s. of (\ref{chieq}) is negative, and this
should remain true for large $J$. The value of the r.h.s. of
(\ref{chieq}) at $\chi=0$ is $\kappa_{0}\le 0$. These two facts
together imply that there's a negative root $\cneg$. Note also that
the value of the r.h.s. of (\ref{chieq}) at $\chi=1$ is
$-v^{2}A_{1}^{2}<0$. This, together with the previous observation,
implies that the two other roots of the polynomial either are both
$<0$ or both $\in [0,1)$ or both $>1$. We're interested in the case
when they both lie in $[0,1)$. We consider ($\cneg,\cmin,\cmax$) as
independent variables that, together with all the previous variables
($\nu, \omega, \upsilon, A_{2}$), satisfy the following conditions
which simply mean that ($\cneg,\cmin,\cmax$) are actually solutions
of the cubic equation:
\begin{eqnarray}\label{cub1}
  \cneg+\cmin+\cmax &=& -\frac{\kappa_{2}}{\kappa_{3}} \\ \label{cub2}
  \cneg\cmin+\cmin\cmax+\cneg\cmax &=& \frac{\kappa_{1}}{\kappa_{3}} \\ \label{cub3}
  \cneg\cmin\cmax &=& -\frac{\kappa_{0}}{\kappa_{3}}.
\end{eqnarray}
We now switch to more convenient variables
$(\widetilde{v},\epsilon)$ instead of $\cmin, \cmax$ (leaving
$\cneg$ unaltered). These two sets are connected in the following
way\footnote{The purpose of introducing the variable $\epsilon$
should be clear --- then the moduli of all tori in our expressions
become $1-\epsilon$. The purpose of introducing $\widetilde{v}$ is
the following: the first parameter of the $\Pi$-function in
(\ref{integrals}) becomes $\frac{\widetilde{v}^2
-1}{\widetilde{v}^2} (1-\epsilon)$, so that it is in direct
correspondence with an analogous parameter in the work \cite{AFZ}.}:
\begin{eqnarray}
\label{newvar}
  \epsilon &=& \frac{\cmin-\cneg}{\cmax-\cneg}\,; \\
  \widetilde{v}^2 &=& \frac{1-\cmax}{1-\cneg}\,; \\
  \cneg &=& \cneg\,.
\end{eqnarray}
Next we write the expressions for all integrals entering our
equations:
\begin{eqnarray}
\nonumber &&\int\limits_{\cmin}^{\cmax}\frac{d\chi}{\chi|\chi'|} =  \frac{2\kappa}{(1-\widetilde{v}^2)^{3/2} (1-\cneg)^{1/2} (1+\cneg\frac{\widetilde{v}^2}{1-\widetilde{v}^2})} \;\Pi\left(\frac{1-\cneg}{1+\cneg\frac{\widetilde{v}^2}{1-\widetilde{v}^2}} (1-\epsilon);1-\epsilon\right) ; \\
\label{integrals}
&&\int\limits_{\cmin}^{\cmax}\frac{d\chi}{(1-\chi)|\chi'|} = -
\frac{2\kappa}{ \widetilde{v}^2 (1-\cneg)^{3/2}
\sqrt{1-\widetilde{v}^2}} \;
\Pi\left(\frac{\widetilde{v}^2-1}{\widetilde{v}^2}
(1-\epsilon);1-\epsilon\right);
\end{eqnarray}
\begin{equation}\nonumber
\int\limits_{\cmin}^{\cmax}\frac{d\chi}{|\chi'|} = \frac{2\kappa\,
K(1-\epsilon)}{\sqrt{(1-\cneg)(1-\widetilde{v}^2)}};
\end{equation}
\begin{equation}\nonumber
\int\limits_{\cmin}^{\cmax}\frac{d\chi\,\chi}{|\chi'|} =
2\kappa\frac{\cneg\, K(1-\epsilon)+ (1-\cneg)(1-\widetilde{v}^2)\,
E(1-\epsilon)}{\sqrt{(1-\cneg)(1-\widetilde{v}^2)}} ;
\end{equation}
\begin{equation}\nonumber
\int\limits_{\cmin}^{\cmax}\frac{d\chi\,(1-\chi)}{|\chi'|} =
-2\kappa\frac{(\cneg-1)\,K(1-\epsilon)+ (1-\cneg)(1-\widetilde{v}^2)
\,E(1-\epsilon)}{\sqrt{(1-\cneg)(1-\widetilde{v}^2)}} .
\end{equation}
Thus, we have chosen the parameter $\epsilon$ rather than $J$ as our
expansion parameter. This means that we have to make an expansion of
the system of equations (\ref{int1})-(\ref{int5}) in $\epsilon$
and determine the corresponding coefficients in the expansion of
various parameters, comparing powers of $\epsilon$ and/or
$\log{\epsilon}$ which arise in this expansion. First of all, before
solving the equations, we get rid of the variable $r$  by plugging the expression for $r$ from
(\ref{int5}) into all other equations.

We make the following ansatz for our parameters:
\begin{eqnarray}
\nonumber
  v(\epsilon) &=& v_{0}(\epsilon)+v_{1}(\epsilon) \epsilon + O(\epsilon^2);
  \\ \nonumber
  \widetilde{v}(\epsilon) &=& \widetilde{v}_{0}(\epsilon)+\widetilde{v}_{1}(\epsilon) \epsilon + O(\epsilon^2);
  \\ \nonumber
  \omega(\epsilon) &=& \omega_{0}(\epsilon)+\omega_{1}(\epsilon) \epsilon + O(\epsilon^2);
  \\ \label{ansatz}
  \nu(\epsilon) &=& \nu_{1}(\epsilon) \epsilon + O(\epsilon^2);
  \\ \nonumber
  A_{1}(\epsilon) &=& A_{1,0}(\epsilon)+A_{1,1}(\epsilon) \epsilon + O(\epsilon^2);
  \\ \nonumber
  A_{2}(\epsilon) &=& A_{2,1}(\epsilon)\epsilon+ O(\epsilon^2);
  \\ \nonumber
  \cneg(\epsilon) &=& \chi_{1}(\epsilon)\epsilon+  O(\epsilon^2);\\ \nonumber
\mathcal{J}(\epsilon) &=&
\mathcal{J}_{0}(\epsilon)+\mathcal{J}_{1}(\epsilon) \epsilon +
O(\epsilon^2),
\end{eqnarray}
where we assume that all "coefficient" functions like
$v_{0}(\epsilon), v_{1}(\epsilon), \widetilde{v}_{0}(\epsilon),$
etc. are terminating series in $\log{\epsilon}$ (this is the reason
why expansions (\ref{ansatz}) are justified). This assumption will
be proved aposteriori --- by the solution that we will find.

We substitute (\ref{ansatz}) into our equations and expand these
equations in $\epsilon^{m}$, ignoring terms with logarithms (that
is, treating any combination $\left(\sum\limits_{k=0}^{n}\,a_{k}
(\log{\epsilon})^k\right) \epsilon^m$ as just $\epsilon^{m}$). Then
we obtain a system of equations for our "coefficient" functions,
which, when solved, exhibits the property of these functions
mentioned above --- that is, they're terminating series in powers of
$\log{\epsilon}$.

In the course of expanding the above written equations we need an
expansion for $\Pi(1-\alpha \,\epsilon,1-\epsilon)$ as $\epsilon \to
0$ ($\alpha$ fixed and $0<\a<1$). To find such an expansion we make
use of the following textbook identity for elliptic functions:
\begin{equation}\label{identity}
\Pi(1-\alpha \,\epsilon,1-\epsilon) = \frac{1}{\alpha (\alpha-1)
\epsilon} \left[\alpha(1-\epsilon) K(1-\epsilon)-(1-\alpha \epsilon)
\Pi(\frac{\alpha-1}{\alpha};1-\epsilon) \right].
\end{equation}
The meaning of using this identity is that it explicitly singles out
the $\frac{1}{\epsilon}$ factor in the expansion. Once we have
written $\Pi(1-\alpha \,\epsilon,1-\epsilon)$ in this form, we may
use \texttt{Mathematica} to generate the expansions of functions in
the r.h.s. of (\ref{identity}):
\begin{eqnarray}\label{expan1}
&&\Pi(1-\alpha \,\epsilon,1-\epsilon)=
\frac{\arctan\left(\sqrt{\frac{1}{\alpha}-1}\right)}{\sqrt
   {\frac{1}{\alpha}-1} \,\alpha\,\epsilon}+\\ \nonumber &+& \frac{\left(2 \alpha
   \sqrt{\frac{1}{\alpha}-1} \arctan\left(\sqrt{\frac{1}{\alpha}-1}\right)+(\alpha-1) (-\log (\epsilon/16)+1)\right)}{4
   (\alpha-1)}+\\ \nonumber&+&\frac{\left(8  \alpha^2
   \sqrt{\frac{1}{\alpha}-1} \arctan\left(\sqrt{\frac{1}{\alpha}-1}\right)
  -(\alpha-1) (2
   \alpha+2 (2 \alpha+1) \log (\epsilon/16)+3)\right) \epsilon}{64
   (\alpha-1)}+O\left(\epsilon^2\right)
\end{eqnarray}
However, in our case $\alpha$ is not constant in $\epsilon$ but
rather depends on $\epsilon$ in the following way:
\begin{equation}\label{alpha}
\alpha(\epsilon)=\frac{\frac{\cneg(\epsilon)}{\epsilon}+(1-\cneg(\epsilon))(1-\widetilde{v}^2(\epsilon))}{1-\widetilde{v}^2
(\epsilon) (1-\cneg(\epsilon))}.
\end{equation}
According to our ansatz (\ref{ansatz}) $\alpha(\epsilon)$ has a
finite positive limit smaller than 1 as $\epsilon \to 0$ --- this is the only thing, which
is important for our expansions to be justified. That is, we plug
the expansion of $\alpha$ in (powers and logarithms of) $\epsilon$
into the expansion for $\Pi(1-\alpha \,\epsilon,1-\epsilon)$
obtained at fixed $\alpha$.

We also need to know the expansion of
$\Pi\left(\frac{\widetilde{v}^2-1}{\widetilde{v}^2}
(1-\epsilon);1-\epsilon\right)$ as $\epsilon \to 0$. It was
constructed in the appendix of \cite{AFZ}. One has to use the
identity
\begin{eqnarray}
&& \Pi\left(\frac{v^2-1}{v^2}
(1-\epsilon);1-\epsilon\right)=\\
\nonumber &=&\frac{1}{\left(1-\left(1-v^2\right) \epsilon
   \right) K(\epsilon )} \left[\frac{1}{2} \pi  v
\sqrt{\left(1-v^2\right)
   \left(1-\left(1-v^2\right) \epsilon \right)}
   F\left(\arcsin\left(\sqrt{1-v^2}\right) ; \epsilon
   \right)+\right.\\ \nonumber &+& \left. K(1-\epsilon )
   \left(\left(1-\left(1-v^2\right) \epsilon \right)
   K(\epsilon )-\left(1-v^2\right) (1-\epsilon ) \Pi
   \left(\frac{v^2 \epsilon }{1-\left(1-v^2\right)
   \epsilon } ; \epsilon
   \right)\right)\right]
\end{eqnarray}
In the r.h.s. there's only one function, which has an expansion that
cannot be directly obtained by \texttt{Mathematica}, and its
expansion looks as follows:
\begin{eqnarray}\nonumber
&& \Pi
   \left(\frac{v^2 \epsilon }{1-\left(1-v^2\right)
   \epsilon } ; \epsilon
   \right) = \frac{\pi
   }{2} + \frac{1}{8} \left(2
   \pi  v^2+\pi \right) \epsilon + \frac{1}{128} \pi  \left(-8
   v^4+44 v^2+9\right) \epsilon ^2+ \\ &+&\frac{1}{512} \pi  \left(16 v^6-72 v^4+206
   v^2+25\right) \epsilon ^3 +O\left(\epsilon^4\right).
\end{eqnarray}
Inverting the expansion
\begin{equation}\label{JJ}
J(\epsilon) = J_{0}(\epsilon)+J_{1}(\epsilon) \epsilon +
o(\epsilon),
\end{equation}
we obtain $\epsilon$ as a function of $J$, that is we return to our
original expansion in the limit $J\to\infty$:
\begin{equation}\label{eps}
\epsilon(J)=\frac{16}{e^{2}} e^{-\frac{\mathcal{J}}{\sin{p\ov 2} }}
\left[1-\frac{8}{e^{2}} e^{-\frac{\mathcal{J}}{\sin{p\ov 2} }} \left( 1-
\mathcal{J}\,\frac{2-3 \sin^{2}{p\ov 2} }{2\sin{{p\ov 2} }} \cos{(\Phi)} -\frac{1}{2}\mathcal{J}^{2}\, \cot^{2}{p\ov 2} \cos{\Phi}
\right)+...\right].
\end{equation}
We now write out explicitly the expansions of the parameters
entering the equations of motion:
\begin{eqnarray}
\label{param}
  \cneg(\mathcal{J}) &=& -\frac{16}{e^2} \sin^{2}{p\over 2}\, \sin^{2}{\Phi \over 2}\, e^{-\frac{\mathcal{J}}{\sin(p/2)}}+...\, , \\ \nonumber
  \cmax(\mathcal{J}) &=& \sin^{2}{p\over 2}\,+ \frac{8}{e^2} \sin{p\over 2}\, \cos^{2}{p\over 2}\; \cos{\Phi} \;(3\,\sin{p\over 2}+\mathcal{J}) \; e^{-\frac{\mathcal{J}}{\sin(p/2)}} + ... \, ,\\ \nonumber
  \cmin(\mathcal{J}) &=& \frac{16}{e^{2}} \sin^{2}{p \over 2}\,
  \cos^{2}\frac{\Phi}{2}\,
  e^{-\frac{\mathcal{J}}{\sin(p/2)}}+ ...\,, \\ \nonumber
  v(\mathcal{J}) &=& \cos{p\over 2}-\frac{4}{e^2} \sin{p\over
  2}\,\cos{p\over
  2}\,\cos{\Phi}\;(\sin{p\over 2}+\mathcal{J})\;e^{-\frac{\mathcal{J}}{\sin(p/2)}}+...\,,\\
  \nonumber \omega(\mathcal{J}) &=& 1 +\frac{8}{e^2} \sin^2{p\over
  2} \, \cos{\Phi} \,e^{-\frac{\mathcal{J}}{\sin(p/2)}} +...\, , \\
  \nonumber \nu(\mathcal{J}) &=& \frac{4}{e^2}
  \cos{p\over 2}\,\sin{\Phi}\,(2\sin{p\over 2}+\mathcal{J})\;e^{-\frac{\mathcal{J}}{\sin(p/2)}}+...\,
  ,\\
  \nonumber f_{0}(\mathcal{J})&=&-\frac{p}{\mathcal{E}}-\frac{\cos{p\over 2}}{\sin^{2}{p\over 2}}+
  \frac{\cos \Phi\, \sin p \,\left(2 \mathcal{J} \cos{p}\,+6 \mathcal{J}-\sin{p\over 2}\,+3 \sin{\frac{3
   p}{2}}\right)}{2 e^{2}\,\sin^{4}{p\over 2}}\,e^{-\frac{\mathcal{J}}{\sin(p/2)}} +...\, ,\\
   \nonumber f_{1}(\mathcal{J})&=& \frac{\cos{p\over 2}}{\sin^{2}{p\over 2}}+\frac{\cos{\Phi}\,\sin{p}\, \left(\sin{\frac{3
   p}{2}}-2 \mathcal{J} (\cos{p}\,+3)-11\sin{\frac{p}{2}}\right)}{2e^2\,\sin^{4}{p\over 2}}  \, e^{-\frac{\mathcal{J}}{\sin(p/2)}} +...\, , \\
   \nonumber a_{0}(\mathcal{J}) &=& -\frac{\delta }{\mathcal{E}}-\frac{4}{e^2} \left(\mathcal{J}+2 \sin{\frac{p}{2}}\,\right) \sin{\Phi}\, \cot
   ^2{\frac{p}{2}}\, e^{-\frac{\mathcal{J}}{\sin(p/2)}} +...\, , \\
  \nonumber a_{1}(\mathcal{J})&=&\frac{8}{e^2}\sin{p\over
  2}\,\sin{\Phi}\,e^{-\frac{\mathcal{J}}{\sin(p/2)}}+...\, ,
\end{eqnarray}
where
\begin{equation}\label{phi}
\Phi=\frac{\delta}{2^{3/2} \cos^{3}(\frac{p}{4})},
\end{equation}
and the solution exists for all $p\in [-\pi;\pi]$ (if and) only if
\begin{equation}\label{restriction}
|\delta|=|2\pi(n_{2}-\gamma J)|\le\pi.
\end{equation}

This means that for the undeformed $AdS_{5}\times S^{5}$, that is
$\gamma=0$, the only possible choice is $n_{2}=0$, or $\delta=0$. In
this case all formulas reduce to what was found in
\cite{AFZ}.\newline To obtain the dispersion relation one should
expand (\ref{int5}) with respect to $\epsilon$ and then substitute
the expansion (\ref{eps}) of $\epsilon$ in terms of $J$. The
dispersion relation with the first correction has the following
form:
\begin{eqnarray}\label{disp}
&&E-J=\frac{\sqrt{\lambda}}{\pi} \sin{\frac{p}{2}}\,
\left(1-\frac{4}{e^{2}}\, \sin^{2}\frac{p}{2} \,\cos{\Phi}\,
e^{-\frac{\mathcal{J}}{\sin p/2 }}+...\right);\\\nonumber
&&\Phi=\frac{\delta}{2^{3/2} \cos^{3}\frac{p}{4}};\;\;\;
|\delta|=|2\pi(n_{2}-\gamma J)|\le\pi.
\end{eqnarray}

\end{document}